# Determination of the optimal shell thickness for self-catalysed GaAs/AlGaAs core-shell nanowires

R. Songmuang*[1, 2], Le Thuy Thanh Giang[1, 2], J. Bleuse[1, 3], M. Den Hertog[1, 2], Le Si Dang[1, 2], and H. Mariette[1,2]

[1]Université Grenoble Alpes, F-38000, Grenoble, France

[2]CNRS, Institut Néel, Nanophysique et Semiconducteurs group, F-38000, Grenoble, France

[3]CEA, INAC-SP2M, Nanophysique et Semiconducteurs group, F-38000, Grenoble, France

ABSTRACT: We present a set of experimental results identifying various effects that govern the carrier dynamics of self-catalyzed GaAs/AlGaAs core-shell nanowires (NWs) grown by molecular beam epitaxy i.e. surface recombination velocity, surface charge traps, and structural defects. Time-resolved photoluminescence of NW ensemble and spatially-resolved cathodoluminescence of single NWs reveal that emission intensity, decay time and carrier diffusion length of the GaAs NW cores strongly depend on AlGaAs shell thickness but in a non-monotonic fashion. Although 7 nm-AlGaAs shell can efficiently suppress the surface recombination velocity of the GaAs NW cores, the effect of the band bending caused by the surface charges remains dominant if the shell thickness is less than 50 nm; that is, the carrier diffusion length is smaller in the NWs with a thinner shell caused by a stronger carrier scattering at the core/shell interface. If the AlGaAs shell thickness is larger than 50 nm, the luminescence efficiency of the GaAs NW cores starts to be deteriorated, ascribed to the defect formation inside the AlGaAs shell evidenced by transmission electron microscopy.

For decades, semiconductor nanowires (NWs) have been regarded as a promising non-planar geometry platform for future nanophotonics and nanoelectronics. Unlike bulk or two-dimensional structures, the NWs are extraordinarily sensitive to surface electronic states owing to a dramatic increasing of their surface-to-volume ratio. In literatures, two major surface effects are generally mentioned: (i) the surface states which act as recombination centers for free carriers[1,2] and (ii) the surface charge traps[3] which induce a pinning of the Fermi level close to the middle of the bandgap, causing the formation of a depletion region near the free surface. The former effect is highly important for the semiconductors with a large surface recombination velocity (SRV) such as GaAs which has SRV as high as $10^6$ cm/s[4,5,6,7].This property is considered as a serious drawback for the NWs based on these materials since their large surface-to-volume ratio strongly enhances such an effect, consequently deteriorating their optical quality e.g. the radiative recombination efficiency[8,9]. The latter effect could either reduce an electrically conducting area of the NWs to the NW center or to completely deplete the NWs[10,11,12,13], limiting their electrical transport feature which is a crucial function in electronic devices.

One way to suppress the surface effects is to epitaxially passivate the NWs by embedding them into another higher bandgap material shell. Indeed, this so-called radial core−shell NW has also attracted progressive attention because of their extended scaling to non−planar geometry which offers an unconventional way of bandgap engineering for novel device architectures[14,15,16]. In the case of GaAs NWs, *in situ* epitaxial growth of AlGaAs layer before exposing its surface to air[17,18,19,20,21] provides an efficient passivation scheme as it prevents the formation of a poor quality native oxide on the GaAs surface. The AlGaAs shell can improve the carrier confinement in the core, suppressing the probability for non-radiative recombination on the GaAs NW core surface caused by their high SRV value. However, the theoretical studies of the strain state of such a one dimensional (1-D) system show that the strain at the GaAs/AlGaAs interface can be rather high[22] despite a small lattice mismatch between GaAs and AlAs. This is due to a shear stress which is important in this 1-D heterostructure configuration. Therefore, a strain relaxation and defect formation in such a small strain system must be considered. So far, the majority of the reports concerning the GaAs/AlGaAs core–shell NWs have focused on a feasibility of the growth[23,24,25,26], elaborating the detailed structural characteristics[27,28,29,30,31], and/or showing the positive effects of the AlGaAs shell on the optical properties of the GaAs NW core[32,33,34]. On the other hand, fewer studies were dedicated to systematically explore relative contributions of various surface effects on the electronic properties of the core-shell NWs, which could have large implications on the optimum design for ultimate performance of the devices based on such complex structures.

Here, we present several experimental results, evidencing an interplay of different effects mentioned above, on the carrier dynamics of the GaAs NW cores. Especially, our work reveals that the contribution of each effect strongly depends on the AlGaAs shell thickness, leading to an optimal shell thickness range (around 50 nm, in our case) for which the electronic properties of the GaAs NW cores are optimized. If the shell thickness is less than 50 nm, despite fully passivating the GaAs NW core free surface, the effect of the band bending induced by the charge carriers available on the outer surface of the core-shell NWs is dominant. This effect alters the scattering process of the excess carriers generated inside the GaAs NW core at the core/shell interface. If the AlGaAs shell thickness is larger than 50 nm, the optical quality of the GaAs NW core starts to be deteriorated, possibly ascribed to the defect formation inside the AlGaAs shell. This non-monotonic influence of the shell thickness on the electronic properties of the core is observed in various optical characteristics of the NW ensemble and single NWs such as PL intensity, energy positions of the emission peak, decay times, and carrier diffusion lengths.

In our experiments, Ga-assisted GaAs NWs were grown on $n$-Si(111) substrates that are covered by native oxide using solid source MBE Riber32P equipped with a conventional As$_4$ source[35]. In order to induce vapor-liquid-solid (VLS) growth of GaAs NWs via Ga droplet catalysts[36,37], Ga and As fluxes were simultaneously deposited on the substrate surface with the deposition rate of 0.1 and 0.15 ML/s, respectively. The GaAs NWs were grown for 90 min, allowing the formation of the GaAs NWs with ~2 μm-length and ~80 nm in diameter. Then, a 10-min growth interruption under As$_4$ flux was applied to completely consume the Ga droplets on the NW top which consequently suppresses the VLS growth mechanism. Hence, the axial growth rate was strongly reduced, leading to an increase of a radial-to-axial growth ratio. The AlGaAs shell growth was performed under As-rich atmosphere with Ga, Al and As fluxes equal to 0.2 ML/s 0.1 ML/s, and 1.5 ML/s, respectively. From the Al/(Ga+Al) flux ratio, the nominal Al content is approximately 33%. The time for the shell deposition was varied from 0-4 hours[38]. Next, a GaAs cap layer with a nominal thickness of 5 nm along the wire radius was deposited to prevent the oxidation of the AlGaAs shell. Note that the Ga, Al and As fluxes were calibrated by using a reflection high energy electron diffraction intensity oscillation during the two-dimensional growth of GaAs and AlAs on GaAs(001) substrate at the same temperature as the NW growth. As deduced from scanning electron microscopy (SEM) images[39], the average AlGaAs shell thickness ($d_{shell}$) increases from 0 to 175 nm as a function of the deposition time. The lateral growth was measured to be ~6 times less than the vertical one. The schematic of the NW and a typical SEM image of the as-grown GaAs/AlGaAs core-shell NW ensemble on Si(111) substrate are presented in Figs. 1(a)-(b), respectively. The SEM image shows the formation of a rough AlGaAs layer in between each core-shell NWs.

Detailed structural characterizations were performed by using high angle annular dark field scanning transmission electron microscopy (HAADF-STEM) on a probe corrected FEI Titan at 300 kV and a probe corrected Titan Ultimate at 200 kV. For luminescence investigations of the NW ensemble, time-resolved photoluminescence (TRPL) was done at 5 K using a frequency-tripled Ti: sapphire laser at a power of 26 μW, [λ = 710 nm (the excitation energy, $E_{ex}$= 1.55 eV) and 800 nm ($E_{ex}$= 1.75 eV), a pulse width ~200 fs, a repetition rate = 76 MHz].  The luminescence was detected by a streak-camera. The effective system resolution is of about 5 ps. Cathodoluminescence (CL) measurements were performed at 7 K in a FEI quanta 200 CL system, using an acceleration voltage of 10 kV and a current of ~30 pA. An electron beam was spatially excited perpendicular to the axis of the NWs dispersed on Si substrates. The overall sample emission was collected by a parabolic mirror, which focused it onto a CCD camera.

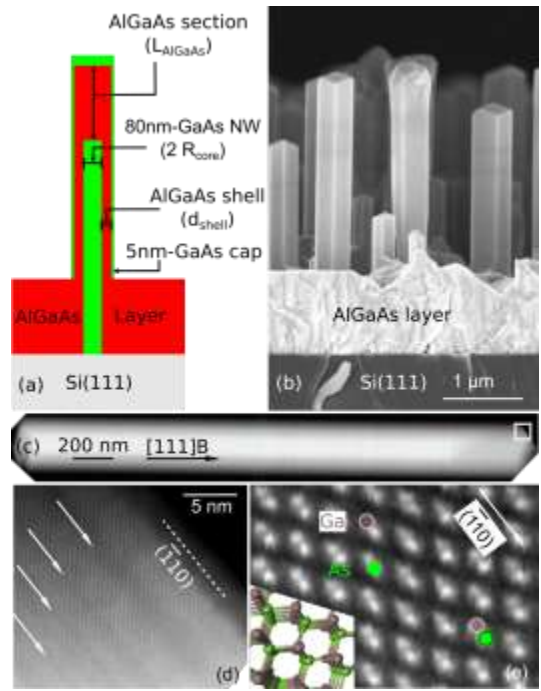

**Figure 1.** (a) Schematic and (b) typical SEM image of as-grown GaAs/AlGaAs core-shell NW ensemble with a 175-nm AlGaAs shell. (c) HAADF-STEM image of a single GaAs/AlGaAs core-shell NW with a 36-nm AlGaAs shell viewed along the [110] direction (d) HAADF-STEM image of the area defined by the square in (c) revealing a spontaneous ordering in the AlGaAs section on the NW top. The dark Al-rich regions are indicated by arrows. (e) The NW polarity showing that the GaAs NWs grows in the [111]B direction. The inset presents an atomic model of the GaAs crystal structure.

The HAADF-STEM image in Fig.1 (c) shows a dark chemical contrast at the top and the side wall regions of the NW, corresponding to an Al-rich area. The high magnification HAADF-STEM of the area defined by the square of this figure is shown in Fig. 1(d), illustrating the alternating brighter and darker regions which correspond to Ga-rich and Al-rich areas, respectively. This image reveals a spontaneous alloy ordering in the AlGaAs sections which aligns parallel with the (-110) facet. A further zoom-in image in Fig. 1(e) shows the zinc blende lattice viewed along the [110] direction whereas the inset presents an atomic model of the GaAs crystal structure. Both group III (Ga or Al) and V (As) columns are clearly visible in a so-called 'dumbbell' structure: two atomic columns is 0.14 nm apart in projection. As the As atom is heavier than the Ga or Al atom, more electrons are scattered on the annular HAADF detector, resulting in a brighter contrast on the As atomic column than on the group III (Ga or Al) column. Hence, Figure 1(e) allows a direct identification of the NW growth direction to be the As polar or [111]B[40].

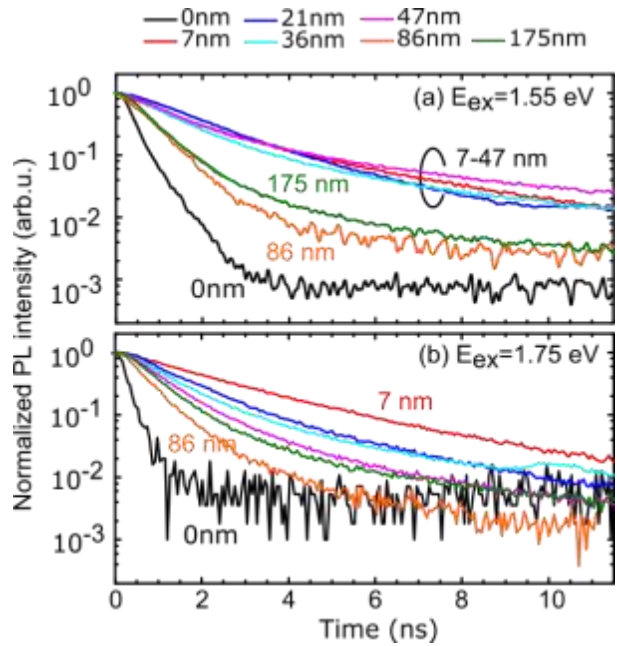

**Figure 2.** Time-resolved spectroscopy of the GaAs NBE emission from the ensemble of GaAs/AlGaAs core-shell NWs with various shell thicknesses, measured by using $E_{ex}$ (a) 1.55 eV and (b) 1.75 eV.

We systematically explored the effect of the AlGaAs shell thickness on the electronic properties of the GaAs NW cores by probing their luminescence characteristics. Figures 2(a)-(b) are time-resolved spectroscopy results showing the decay of the normalized PL intensity of the near band edge (NBE) emission of the GaAs NW cores with various AlGaAs shell thicknesses as a function of time measured by using $E_{ex}$=1.55 eV and 1.75 eV, respectively. Clearly, the decay behavior of PL intensity is drastically different between two values of $E_{ex}$, which we will discuss later in the next paragraphs.

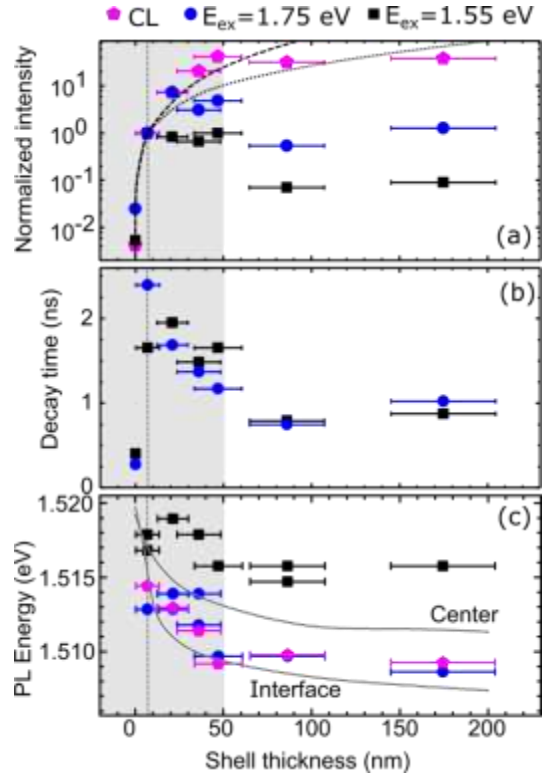

**Figure 3.** (a) Normalized integrated PL and CL intensities (b) decay times, and (c) PL and CL peak energies plotted as a function of AlGaAs shell thicknesses. The PL is measured by using the exciting photon with $E_{ex}$=1.55 eV and 1.75 eV. The dashed and dotted lines in (a) correspond to the values derived from a simple model assuming that the increasing intensity is coming from the excess carriers travelling from the AlGaAs section on the NW top and from the AlGaAs shell, respectively. The grey solid lines in (c) present the lowest energy distance between the bottom of conduction band to the top of valence band calculated at the center of the GaAs NW core and the GaAs/AlGaAs NW sidewall interface. All the measurements were performed at 7K.

Figures 3(a)-(c) summarize the normalized integrated luminescence (CL and PL) intensities, the energy positions of the emission peak and its decay times of the GaAs NBE emission from the GaAs NW cores as a function of the AlGaAs shell thickness ($d_{shell}$), respectively. As the decay curves in Figs. 2(a)-(b) cannot be fitted with a single exponential decay, we define the decay time ($\tau_d$) as the time for which the maximum intensity decreases by a factor of $1/e$, for data comparison. We can separate the experimental data in Fig. 3 into 2 regimes corresponding to the shell thickness lower (grey area) and higher than 50 nm (unshaded area). The separation between the two different regimes is defined at the point where the integrated PL intensity of the GaAs NW core decreases by one order of magnitude while the integrated CL intensity becomes constant.

In the first regime ($d_{shell} < 50$ nm), a drastic increase of the integrated intensity of GaAs NW core emission more than two orders of magnitude was observed after a 7 nm-AlGaAs shell thickness was deposited, at any excitation method (i.e. PL with $E_{ex}$=1.55 eV and 1.75 eV as well as CL). In the case of $E_{ex}$= 1.55 eV (black square symbol), the PL intensity remains constant if the shell thickness is in the range of 7 -50 nm. These results reveal that by using this excitation energy, the excess carriers were generated only inside the GaAs NW core and that the 7-nm AlGaAs shell is *sufficient* to suppress surface recombination processes at the GaAs NW core surface. For this reason, both PL and CL intensities are normalized with the luminescence intensity from the GaAs NW core with the 7-nm AlGaAs shell. Differently, the PL intensity measured by using $E_{ex}$=1.75 eV (blue circle symbol) continuously increases until the AlGaAs shell thickness reaches 21 nm. Although the value of this $E_{ex}$ is lower than the nominal energy gap of $Al_{0.33}Ga_{0.67}As$ which is ~1.9 eV[41], a weak PL intensity locating at the energy above the GaAs bandgap but lower than 1.9 eV was observed. This could be explained by the energy fluctuations induced by spontaneous alloy ordering, alloy fluctuations [28,29,30,31] and/or point defects in the shell[42] , allowing the $Al_{0.33}Ga_{0.67}As$ area to be *partially* excited. The slightly higher PL intensity of the GaAs NW core in comparison to the one obtained by $E_{ex}$= 1.55 eV is attributed to an increasing number of photo-generated carriers transferred from the shell to the core which then recombine radiatively.

For the CL measurements, we used the high excitation energy of the electron beam of 10 kV which corresponds to an electron penetration depth of ~0.5 μm[43,44], allowing the $Al_{0.33}Ga_{0.67}As$ section to be *fully* excited in the first regime. Therefore, a larger number of excess carriers can be transferred from the $Al_{0.33}Ga_{0.67}As$ to the GaAs NW core in comparison to the one in the PL. This fact increases the CL intensity even further until the shell thickness reaches 50 nm. To qualitatively describe the CL intensity evolution as a function of the shell thickness in the range of 0-50 nm, we consider three regions where the excess carriers are created i.e. the GaAs NW core, the AlGaAs shell and the AlGaAs section on the top of GaAs NW [see Fig. 1(a)]. As a first approximation, we assume that the luminescence intensity is directly related to the number of generated electron-hole pairs which is proportional to the excited volume. Such an assumption is valid when the non-radiative recombination processes at the NW surface were completely suppressed. If the carriers were transferred only from the AlGaAs section on the NW top, the CL intensity, $I(d_{shell})$, of the GaAs NW core could be written as

$$I(d_{shell}) \propto (R_{core}+d_{shell})^2 L_{AlGaAs} \qquad (1)$$

, where $d_{shell}$ is the AlGaAs shell thickness, $R_{core}$ is the GaAs NW core radius, and $L_{AlGaAs}$ is the axial length of the AlGaAs section which is 6 times of $d_{shell}$. On the other hand, if the excess carriers were transferred from the AlGaAs shell area, we can describe

$$I(d_{shell}) \propto R_{NW}^2 - R_{core}^2 \qquad (2)$$

, where $R_{NW}$ is the total radius of the NW core shell which is equal to $R_{core}+d_{shell}$. The fitting curves using the two assumptions are presented by the dashed and dotted lines in Fig. 3(a), respectively. By comparing with the experimental results, we deduce that most of the excess carriers which radiatively recombine in the GaAs NW core were transferred from the volume of AlGaAs section on the NW top.

In the second regime ($d_{AlGaAs}$ > 50 nm), the integrated PL intensity of the GaAs NW core decreases by one order of magnitude [Fig. 3(a)]. As the luminescence obtained by $E_{ex}$= 1.55 eV is only coming from the recombination of the carriers generated in the GaAs NW core, the decrease of PL intensity is attributed to the deterioration of the material quality of the core or the core/shell interface, possibly due to the defect formation. The CL emission does not monotonically increase because of the limited excited volume in AlGaAs section. Moreover, the slight decreasing of CL intensity might also be explained by the defects which limit the carrier diffusion length in AlGaAs section, thus preventing the excess carriers to reach the GaAs NW core once the shell thickness is above 50 nm. The interpretation is supported by the appearance of the emission at the energy higher than the GaAs energy gap in the CL spectrum (not shown here), indicating that the excess carriers in the AlGaAs sections recombine before reaching the core.

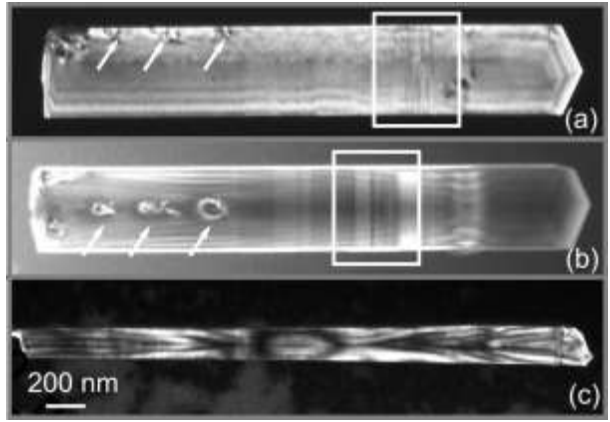

**Figure 4.** presents DF-TEM images of individual NWs with a nominal AlGaAs shell thickness of 175 nm [(a), (b)] and of 21 nm [(c)] obtained on the <220>, <111>, and <220> reflections, respectively, in two-beam condition close to [(a), (b)] or on (c) the [112] ZA.

To assess the presence of the defects appearing in the core-shell NWs, the individual NWs with a shell thickness of 175 nm [Figs. 4(a)-(b)] and of 21 nm [Figs. 4(c)] were oriented in a two-beam condition close to a [112] zone axis (ZA) for dark field (DF)-TEM imaging. These images were made on the <220> or <111> reflections. All images in Fig. 4 show fringes mostly parallel to the NW sidewalls with slowly varying contrasts along the NW length. These are so-called equal thickness fringes caused by constructive or destructive interference of the imaging diffracted beam. The contrast can be varied along the NW axis if the diffraction conditions change; for instance, if the NW was bent.

Diffraction contrast induced by defects can be observed at the side of the NW near the NW base [Fig. 4(a)] and at the center of the NW [Fig. 4(b)], as indicated by the white arrows. When we rotated around the axis of the NW in Fig. 4(b), the defect contrast moves to the side of the NW, evidencing that these defects are indeed present in the AlGaAs shell. Differing from the gradually varying equal thickness fringes, this contrast varies rapidly and is present only at well-defined locations. Therefore, it must be related to crystal defects which are most likely dislocations. On the other hand, such a contrast is not visible in the DF-TEM image of the GaAs NW with a 21-nm shell shown in Fig.4(c). Such defects in a thick AlGaAs shell could possibly extend up to the interface between the GaAs NW core and AlGaAs shell and behave as a nonradiative recombination channel which is responsible for the decreasing of the PL intensity in the second regime ($d_{shell} > 50$ nm). The vertical lines visible in most DF-TEM images [for examples, the area indicated by the white squares in Figs. 4(a)-(b)] are ascribed to twin defects or to hexagonal wurtzite domains in the cubic zinc blende crystal structure which are typically found in Ga-assisted GaAs NWs[45].

Similar to the behavior of the luminescence intensity, the decay time and the spectral peak position also depend on the laser excitation energy, $E_{ex}$. The summary in Fig. 3(b) shows that the decay time increases significantly from 0.3 ns to 1.7 ns ($E_{ex}$= 1.55 eV) or from 0.4 ns to 2.4 ns ($E_{ex}$=1.75 eV) when the 7-nm AlGaAs shell was deposited. This behavior supports the explanation that this shell configuration is *sufficient* to suppress non-radiative recombination processes at the GaAs NW surface. When the shell is between 7-50 nm, the decay time measured by $E_{ex}$=1.55 eV is scattering at around 1.5-1.9 ns [see also Fig. 2(a)] while the ones obtained by $E_{ex}$=1.75 eV clearly decrease from 2.4 ns to 1 ns [see also Fig. 2(b)]. Such an evolution of the decay time as a function of the shell thickness was also observed in the GaAs/AlGaAs core-shell NWs grown by metal organic chemical vapor deposition[30]. Concerning the spectral peak positions, Fig. 3(c) summarizes those of PL and CL spectrum as a function of the shell thickness. In this first regime, with $E_{ex}$=1.55 eV, the PL peak energies of the GaAs NW core slightly decreases from 1.519 eV to 1.516 eV, while the ones obtained by using $E_{ex}$=1.75 eV and from the CL measurements systematically red-shift from 1.515 eV to 1.505 eV, when the shell thickness increases from 7 nm to 50 nm.

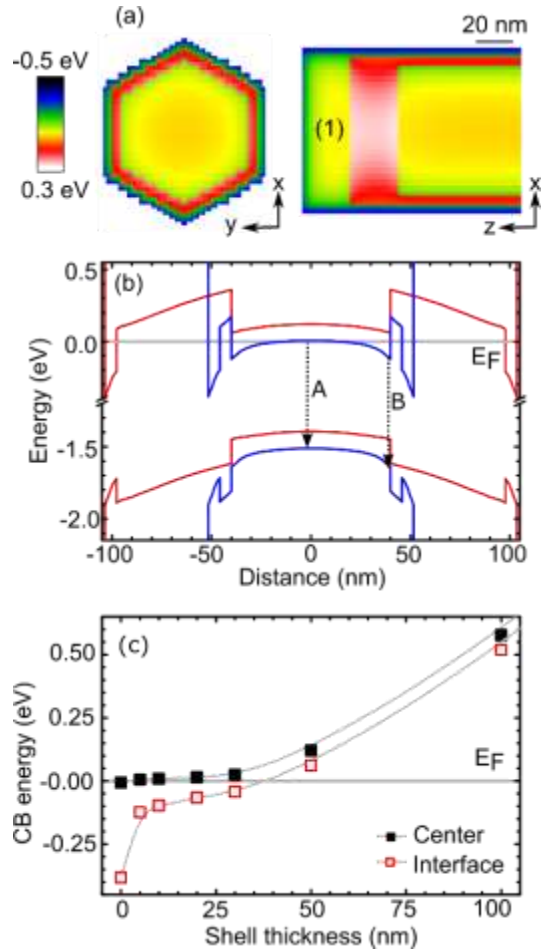

**Figure 5.** (a) Contour plot of the conduction band energy at low temperature (7) K in respect to $E_F$ in *x-y* and *x-z* directions (b) Conduction and valence band energies across the wire diameter taken at the center of the $Al_{0.33}Ga_{0.67}As$ core-shell NW with 5 nm (blue) and 50 nm shell (red) at z = 90 nm from the top $GaAs/Al_{0.33}Ga_{0.67}As$ interface [interface (1)]. (c) CBE energy taken at the GaAs NW center and at the sidewall interface between the GaAs NW core and the $Al_{0.33}Ga_{0.67}As$ shell plotted as a function of the shell thckness. $E_F$ locates at 0 eV.

To understand the evolution of the luminescence (PL and CL) peak energy of the GaAs NW core and the decay time as a function of the shell thickness, we performed 3D-simulations for the bandstructure of a $GaAs/Al_{0.33}Ga_{0.67}As$ core-shell NW by using Nextnano[3] solver.[46], using the structural parameters deduced from our experiments. For the simulations, we define the hexagonal cross-sectional NW with the core diameter of 80 nm. The vertical growth rate of the $Al_{0.33}Ga_{0.67}As$ shell and the GaAs cap is 6 times larger than the lateral ones. Here, the strain minimization model was applied, allowing the strain in the NW to elastically relax in three dimensions. The donor surface state density on the GaAs cap layer surface is set to be $5x10^{12}$ $cm^{-2}$. We assume p-type GaAs NWs with a carbon impurity concentration of $10^{17}$ $cm^{-3}$ which is the background doping density in our growth chamber. The calculation temperature is 7 K.

Figure 5(a) is a contour plot of the conduction band in respect to the Fermi level in *x-y* (top left panel) and *x-z* (right panel) directions. In Fig. 5(b), the conduction and valence bands across the wire diameter, taken at the center of a GaAs/$Al_{0.33}Ga_{0.67}As$ core-shell NW with a 5-nm and 50-nm $Al_{0.33}Ga_{0.67}As$ shell are shown. The conduction band energies taken at the center of the GaAs NW core and at the sidewall GaAs/$Al_{0.33}Ga_{0.67}As$ interface are plotted as a function of the shell thickness in Fig. 5(c). All the represented values were taken at 90 nm away from the GaAs/$Al_{0.33}Ga_{0.67}As$ top interface [interface (1)] since the band energy along the NW axis after this point is constant. Our simulations reveal that within a certain range of shell thickness, the considered charge carrier density on the NW surface is high enough to alter the GaAs NW core from *p*-type to be *n*-type; that is, the Fermi level ($E_F$) is shifted to be close to or higher than the conduction band edge (CBE) of the GaAs core.

Figure 5(c) illustrates that the $E_F$ of the NWs is about in resonance with or *slightly below* the CBE at the NW center, whereas the $E_F$ is lying *above* the CBE at the sidewall interface between the GaAs NW core and the $Al_{0.33}Ga_{0.67}As$ shell when the shell thickness is less than 30 nm. For example, Figure 6(a) presents the band diagram of the GaAs NW with a 20-nm $Al_{0.33}Ga_{0.67}As$ shell, showing that the energy distance between the CBE and the valence band edge (VBE) at the NW center [position A in Fig. 5(b)] is 1.515 eV, while Fig. 6(b) shows that the one between $E_F$ and the VBE at the sidewall interface [position B in Fig. 5(b)] is 1.577 eV. Hence, the photons with $E_{ex}$ 1.55 eV can only excite the NW center. Differently, the photon energy of 1.75 eV and the electron energy of CL are large enough to excite the whole GaAs NW core and the $Al_{0.33}Ga_{0.67}As$ area. In both cases, the energy of the emitting photons ($E_{emit}$) corresponds to the excitonic bandgap which changes slightly from the NW center to the GaAs/$Al_{0.33}Ga_{0.67}As$ sidewall interface. We plot the expected luminescence energy derived from the lowest *e-h* transition energy of the GaAs at the NW center and at the GaAs/$Al_{0.33}Ga_{0.67}As$ sidewall interface as the grey solid lines in Fig. 3(c). The simulations show that the transition energies at the interface red-shifts from the ones taken at the NW center. The comparison between the simulations and the experimental data suggests that the PL obtained by $E_{ex}$ = 1.55 eV is due to a carrier recombination at the center of the GaAs NW, while the CL and the PL with $E_{ex}$ =1.75 eV mostly result from the carriers which recombine near the GaAs/$Al_{0.33}Ga_{0.67}As$ interface[47].

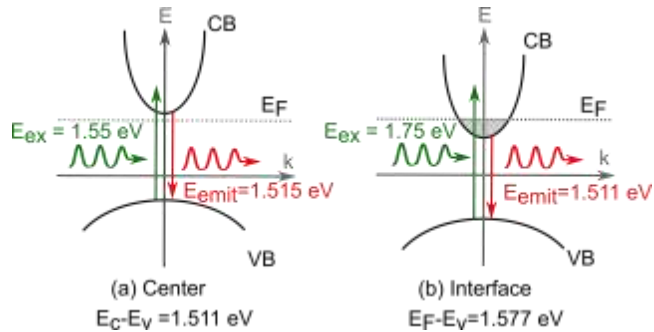

**Figure 6.** Band diagram of the core-shell NW with a 20 nm shell (a) at the GaAs center excited with $E_{ex}$=1.55 eV and (b) at the GaAs/$Al_{0.33}Ga_{0.67}As$ sidewall interface excited with $E_{ex}$ =1.75 eV. Note that the schematic is not to scale.

Such an interpretation is also supported by a monotonically shortening of the decay time measured by using $E_{ex}$ =1.75 eV [see Fig. 2(b)] when the shell thickness increases from 7 nm to 50 nm. This is because the electric field induced by the surface band bending changes the overlap of the wave function of the *e-h* pair, and thus influences on the carrier recombination lifetime. For the larger shell thickness, the effect of the electric field induced by the surface charges is less significant, resulting in a shorter decay time for the NW with a thicker shell. Differently, a carrier recombination at the center area of the GaAs NW core is less sensitive to this electric field, leading to a constant decay time in the PL with $E_{ex}$=1.55 eV [Fig. 2(a)]. The small red energy shift in this case is mainly caused by the residual tensile strain induced by the $Al_{0.33}Ga_{0.67}As$ shell[48].

The surface charge induced band bending not only affects the NW luminescence characteristic but it also plays a role on the carrier transport inside the GaAs NW core. To evidence such an effect, we performed spatially-resolved CL measurements of several individual NWs at 7 K (~30 wires). Figures 7 (a) and (b) show the schematic of a single GaAs/AlGaAs core-shell NW and the SEM of a typical dispersed GaAs NW with an 85-nm shell. The normalized integrated CL intensity variation of the GaAs NW core along the wire axis presented in Fig. 7(c) is corresponding to the NW shown in Fig. 7(b).

As a general trend, when the exciting electron beam is moved from the top to the base of the NW, the CL intensity of GaAs monotonically increases to the maximum value and then decreases. Figure 7(c) shows that the maximum value is at around 700 nm from the NW top, fairly agreeing with the expected length of AlGaAs section on the top of this particular wire (~500 nm). When the electron beam was located on the NW top, the excess carriers that were generated in the AlGaAs can travel to the GaAs NW core and then radiatively recombined. As a simple approximation, we can deduce the *effective* carrier diffusion length in the AlGaAs section ($\lambda_{AlGaAs}$) from the luminescence intensity variation of the GaAs NBE emission by

$$I(z)=I_0 \exp((z-L_{AlGaAs})/\lambda_{AlGaAs}), \ 0 \leq z \leq L_{AlGaAs} \qquad (3)$$

, where z is the distance along the wire axis from the excited area to the second AlGaAs/GaAs interface [interface (2)] and $I_0$ is the intensity equivalent to the generated carriers at the excited volume. The fitting curve is shown by the blue solid line in Fig. 7(c). Since the measurements were performed at low temperature, the extracted $\lambda_{AlGaAs}$ corresponds to the effective diffusion length of the exciton in AlGaAs, which fluctuates between 50 - 300 nm.

Once the exciting electron beam entered the GaAs NW core region, the luminescence of the GaAs NBE is mostly from the recombination of the excess carriers that are generated inside the GaAs core rather than the one in the AlGaAs shell. Indeed, the bottom surface of the dispersed NWs is a free surface since it is not passivated by the AlGaAs shell. Thus, the emission intensity of the GaAs NW core is sensitive to the nonradiative recombination process at that region. The intensity evolution in the other part of the data of Fig. 7(c) starting from the position which show the maximum intensity to the bottom part of the NW can be described by

$$I(z)=I_0(1-\exp(z/\lambda_{GaAs})), \ L_{AlGaAs}<z< L \qquad (4)$$

, where $\lambda_{GaAs}$ is the carrier diffusion length along the axis of GaAs NW core. L is the distance from the first GaAs/AlGaAs interface [interface (1)] to the bottom of the NW (the NW/air interface). The fitting curve is represented by the red solid line.

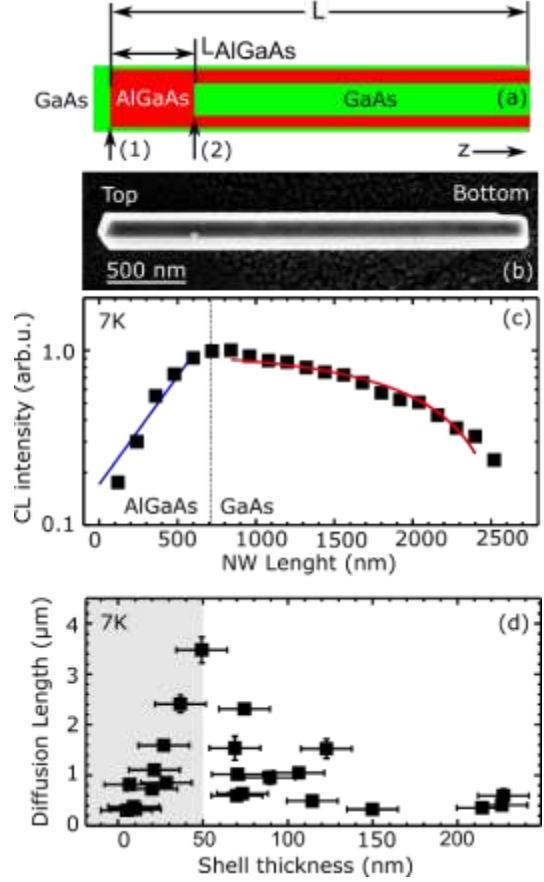

**Figure 7.** (a) Schematic of a single GaAs/AlGaAs core-shell NW (b) SEM image of a typical dispersed GaAs NW with an 85 nm shell. Note that the schematic in (a) is not to scale. (c) Normalized integrated luminescence intensity of GaAs as a function of a distance along the wire axis measured from the NW shown in (b). Blue and red solid lines are the fitting to extract the carrier diffusion length in the AlGaAs sections and the GaAs NW cores, respectively. (d) Summary of the carrier diffusion lengths in the GaAs NW cores as a function of shell thickness at 7 K.

The extracted $\lambda_{GaAs}$ as a function of the shell thickness is summarized in Fig. 7(d). Our results reveal that $\lambda_{GaAs}$ increases from 0.2 μm to 4 μm when the AlGaAs thickness increases from 7 nm to 50 nm. Although the 7-nm AlGaAs shell thickness is *sufficient* to passivate the GaAs NW core surface for a nonradiative recombination process, this value is not enough to improve the carrier mobility inside the GaAs NW core. We can clearly identify the experimental results into 2 regimes similar to the ones shown in Fig. 3. In the first regime where $d_{shell} < 50$ nm represented by the grey region of Fig. 7(d), the surface induced band bending enhances the localization of the electrons to the area close to the GaAs/AlGaAs interfaces, thus increasing a scattering probability. This effect is stronger in the NWs with a thinner shell which results in a smaller carrier diffusion length. The effect of the surface charges should be suppressed when $d_{shell} > 50$ nm or the unshaded region of Fig. 7(d). We should expect a constant $\lambda_{GaAs}$ but a systematic decrease of this value was found. We attribute this counterpart effect to the defects in the AlGaAs shell which might propagate to the GaAs /AlGaAs interface and influence the optical properties of the core. This interpretation is corroborated by the reduction of the emission intensity [(Fig 3(a)] and the decrease of decay time [Figs. 2(b) and 3(c)] from the core-shell NWs with $d_{shell} > 50$.

In conclusion, we show that the optical properties of GaAs/AlGaAs core-shell NWs are strongly governed by the interplay between three effects, namely (i) the surface recombination velocity, (ii) the surface charge traps, and (iii) the structural defects in the AlGaAs shell. While the first effect can be easily suppressed by passivating the NW surface with a thin shell layer (7 nm), the second one requires a thicker shell (~50 nm) to be eliminated. However, further increasing the shell thickness leads to the formation of structural defects in the AlGaAs shell (the third effect) which strongly limits the optical quality and carrier dynamics in the GaAs NW core. With systematic and quantitative analyses of all these effect, a trade-off required to optimize the emission intensity, decay time and diffusion length in the GaAs NW core, is the AlGaAs shell thickness of 50 nm.

AUTHOR INFORMATION

Corresponding Author

*E-mail: rudeesun.songmuang@neel.cnrs.fr

ACKNOWLEDGMENTS

This work was performed in the CEA-CNRS team "Nanophysique and Semiconductors" of INAC and Institute Néel, and in the team "Materials Structure and Radiation" of Institut Néel. We thank for scientific support from F. Donatini for CL measurements and fruitful discussion with R. André and N. T. Pelekanos. We acknowledge the technical support from the "Nanofab" team of Institut Néel (B. Fernandez) and benefit from the technological platform "NanoCarac" of CEA-Minatech. Part of this work is financially supported by ANR program JCJC (Project COSMOS, ANR-12-JS10-0002).